# Thermal Conductivity of Core-Shell Nanocomposites for Enhancing Thermoelectric Performance

S. J. Poon, A. S. Petersen, and Di Wu

Department of Physics, University of Virginia, Charlottesville, VA 22904-4714

Abstract

The differential effective medium method (DEM) is presented from a physical viewpoint and employed to calculate the lattice thermal conductivity of nano-bulk composites comprising of core-shell particles. Extended from the average-T-matrix single-particle approximation, DEM incorporates multiparticle effect essential for the study of core-shell nanocomposites (CSN). Interparticle boundary scattering in addition to intraparticle boundary scattering in CSN is found to add to the reduction of thermal conductivity of nanocomposites. Thus, CSN hold the promise of improving the thermoelectric dimensionless figure of merit ZT above that of monolithic nano-bulk phases. Si and SiGe based CSN serve as illustrative examples.

Key words: effective medium model, core-shell semiconductor, nanocomposite, thermal conductivity.

Thermoelectric materials with a dimensionless figure of merit ZT ($=S^2\sigma T/\kappa$) higher than 1 are desirable for thermal to electric energy conversion, where S is the Seebeck thermopower, $\sigma$ the electrical conductivity, and $\kappa$ the thermal conductivity equal to $\kappa_e + \kappa_L$, sum of the electronic and lattice thermal conductivities. Nanostructuring approach has shown promise for designing high-ZT materials, with ZT reaching as high as 1.8.[1-7] Nanocomposites contain a high density of interfaces that act as both thermal barriers and scatterers of long-wavelength phonons[8-10] as well as energy filters for carriers, resulting in the reduction of thermal conductivity and increase of thermopower.[3, 11] There are currently several *ab-initio* methods in use to determine thermal properties. Several utilize the Boltzmann transport equation to calculate the thermal conductivity of core-shell systems, including Monte Carlo simulations[8] and numerical calculations using either the gray model[12, 13] or the non-gray model.[14, 15] There have also been first-principles based calculations, using a Kubo-Greenwood style approach, accounting for disorder-induced scattering that many models based on the Peierls-Bolztmann equation, coherent potential approximation, or atomic models fail to take into account.[16, 17] Recently, the differential effective medium (DEM) model[18] was employed as an alternative to ab-initio calculation for investigating the thermal conductivity of nanocomposites.[19] Effective medium approaches have been used in electronic structure calculations, they can also be applied to the modeling of thermal transport.[20] These models use scattering off of impurities to determine the properties of the material on average. This same type of analysis can be conducted by scattering phonons off of the impurities in alloys or, in the case of composites, off of the particles to determine average thermal conductivity. However, the average-T-matrix, single-particle approximation (ATA) approach to phonon scattering in core-shell systems, notably those with tubular nanostructures,[21] is inadequate at high particle volume fraction and multiparticle effect must be included. As shown earlier,[19] DEM model inherently contains interparticle phonon scattering effects that are important for the reduction of thermal conductivities in nanocomposites. Introduced as a phenomenological model, the DEM method does require several pre-determined parameters, when compared to ab-initio nearly parameter free methods, the extensive computer simulations required in those computations renders the DEM method as a more convenient path for the analysis of prospective high-ZT systems.

In this paper, the DEM model is presented from a physical perspective to facilitate the investigation of the effect of intergrain boundary phonon scattering in core-shell nanocomposites (CSN). From the thermal stability viewpoint, CSN structure can slow down or suppress grain growth. Significant reduction of thermal conductivities in several CSN materials is observed. Our model CSN is constructed by embedding particles (acting as cores) homogeneously in a host (acting as shell). In practice, the CSN is formed by consolidating or assembling the core-shell particles. At low particle volume fraction, phonon-particle scattering can be described by hard-sphere scattering. On the other hand, intergrain boundary scattering, not considered earlier,[8, 19] is dominant at high particle volume fraction and it must be considered in computing the thermal conductivity of CSN.

In the average-T-matrix approximation,[20, 22] the effective thermal conductivity of an isotropic composite embedded with a small volume fraction $\phi$ of particles is

$$k = k_h \frac{k_p(1+2\alpha) + 2k_h + 2\phi(k_p(1-\alpha) - k_h)}{k_p(1+2\alpha) + 2k_h - \phi(k_p(1-\alpha) - k_h)} \quad (1)$$

where $k_h$ is the thermal conductivity of the host matrix, $k_p$ the thermal conductivity of the particles, $\alpha$ the dimensionless thermal resistance parameter defined as $\alpha = Rk_h/(a/2)$, R is the thermal boundary resistance, and a is the diameter of the particle inclusions. Applying the differential method to equation (1), k is obtained by increasing the volume fraction by $d\phi$ each time. The key idea is that at volume fraction $\phi$, the thermal conductivity of the matrix is updated to $k(\phi)$ of the composite. Previously, $d\phi$ was substituted by an effective $d\phi' = d\phi/(1-\phi)$, where $1-\phi$ is the volume of the unoccupied host.[18-20, 22] There are two issues with this empirical substitution. First, the physical meaning of $d\phi'$ is unclear. Second, that $d\phi'$ becomes very large when $\phi \rightarrow 1$. Alternatively, by formulating the DEM model from a more physical viewpoint, the approach has the advantage of being more tractable for incorporating interparticle scattering effect. Concisely, after adding $d\phi$ of particles to the instantaneous host, the host's conductivity $k_h(\phi)$ prior to considering phonon scattering from the added particles is $k(\phi/(1-d\phi))$. The effective medium's volume fraction $1-d\phi$ does not include $d\phi$ and it is always near 1. Since $k(\phi/(1-d\phi)) \approx k(\phi) + k'(\phi)\phi d\phi$, the transformation relation $k_h(\phi) \rightarrow k(\phi) + \phi dk(\phi)$ is obtained.

Intragrain and intergrain phonon scattering must be considered. $k_p$ is replaced by $k_{p0}/(1+L_p/a)$, where $L_p$ is the intragrain phonon mean free path and $k_{p0}$ is the intrinsic lattice thermal conductivity of particle material in bulk form. $k_h$ is reduced due to phonon-particle scattering. For hard-sphere scattering within ATA, $k_h(\phi) = k_{h0}/(1+3L_h\phi/2a)$, where $\phi$ is small, $2a/3\phi$ is the phonon-particle scattering mean free path in the host,[8] and $k_{h0}$ is the intrinsic lattice thermal conductivity of host material in bulk form. As demonstrated earlier, this mean free path essentially applies to different space-filling polyhedral particles with the geometric mean of their principal dimensions represented by a.[19] Thus, hereafter the term "hard-sphere scattering" will be replaced by "hard-particle scattering". In DEM, $k_h(\phi+d\phi)=k_h(\phi)/(1+3L_h(\phi)d\phi/2a)$, which can be rewritten as $k_h(\phi)/(1+(3L_h(\phi)\phi_f/2a)(d\phi/\phi_f))$, where $L_h(\phi)$ is the host's phonon mean free path, and $\phi_f$ is the target filling fraction. As shown below, at high $\phi$ the hard-particle scattering model employed in previous work[8, 19] will be replaced by one that prescribes boundary scattering between neighboring particles. $k_h(\phi)$ is written as $k_h(\phi)=k_h/(1+L_hF(\phi)/a)$, where $F(\phi)$ is a function of the interparticle distance. Following the hard-particle case, the particles configuration at $\phi_f$ is used as a template for adding particles starting from $\phi=0$. That is, the increment in phonon-particle scattering scales with $d\phi$, and $k_h(\phi+d\phi)$ is given by:

$$k_h(\phi+d\phi) \to \frac{k_h(\phi)}{1+\left(\frac{L_h(\phi)F(\phi_f)}{a}\right)\frac{d\phi}{\phi_f}} \qquad (2)$$

Substituting $k_h(\phi)$ in (2) and keeping only the first-order terms, $k_h(\phi+d\phi)$ can be expressed as:

$$k_h(\phi+d\phi) \to k(\phi)+\phi dk(\phi)-k(\phi)\left(\frac{L(\phi)F(\phi_f)}{a\phi_f}\right)d\phi \qquad (3)$$

where $L_h(\phi)$ has been replaced by $L(\phi)$. In equation (1), $k_h$ becomes $k_h(\phi+d\phi)$, and using the transformation relations introduced, $k(\phi+d\phi)$ is given by

$$k(\phi+d\phi)=k(\phi)+\phi dk(\phi)-k(\phi)\left(\frac{L(\phi)F(\phi_f)}{a\phi_f}\right)d\phi+3k(\phi)d\phi\left(\frac{k_p(1-\alpha(\phi))-k(\phi)}{k_p(1+2\alpha(\phi))+2k(\phi)}\right) \qquad (4)$$

After rearranging terms, the following differential form of $k(\phi)$ is obtained:

$$dk(\phi)=\frac{3k(\phi)d\phi}{1-\phi}\left(\left(\frac{k_p(1-\alpha(\phi))-k(\phi)}{k_p(1+2\alpha(\phi))+2k(\phi)}\right)-\frac{L(\phi)F(\phi_f)}{3a\phi_f}\right) \qquad (5)$$

$L(\phi)$ in the equation is $3k(\phi)/C(\phi)v(\phi)$, where C and v are the lattice heat capacity and average phonon velocity, respectively. For hard-particle scattering, $F(\phi)=3\phi/2$ and equation (3) of reference 19 is obtained. It was shown that integration of equation (5) effectively took into account some degrees of multiparticle effect as the particles are added to the host. In the present treatment, interparticle effect is also implicitly included in $F(\phi)$ that describes phonon scattering between the nearest-neighbor particles.

The dependence of $F(\phi)$ on the interparticle distance t, defined here as the shortest distance between the particles or twice the shell thickness, underlies a stronger reduction in thermal conductivity. From $\phi=(t/(t+a))^3$, $t(\phi)$ is given by $t(\phi)=a(1-\phi^{1/3})/\phi^{1/3}$.[19] As shown above, for hard particles, the scattering rate is proportional to $F(\phi)/a \sim \phi/a$. The rate increases with $\phi$ and saturates at $\sim 1/a$. As t decreases, the scattering process is more appropriately described by interparticle grain-boundary scattering. The relevant scattering length changes from a to t at some crossover volume fraction, $\phi_0$. To estimate $\phi_0$, let $t \approx a$, which gives $\phi_o \sim 0.125$. For $\phi > \phi_o$, $1/a < 1/t$ and interparticle boundary scattering becomes important, replacing hard-particle scattering. As $\phi \to 1$, $t(\phi)$ decreases below certain phonon wavelength $\lambda_h$ of the host, intergrain scattering is suppressed and the system essentially becomes a monolithic nano-bulk system. Thus, overall, $F(\phi)$ can be represented by the following empirical equation:

$$F(\phi) = \left(\frac{3}{2}\phi e^{-\phi/\phi_0} + \frac{a}{t(\phi)}(1-e^{-\phi/\phi_0})\right)(1-e^{-(t(\phi)/\lambda_h)^n}) \qquad (6)$$

$F(\phi)$ represented by equation (6) has the prescribed dependence of phonon-particle scattering on $\phi$ (c/o equation (2)) namely: $F(\phi_f)d\phi/a\phi_f \sim 3d\phi/2a$ at low $\phi_f$ (hard particle), $\sim d\phi/t(\phi_f)$ at intermediate $\phi_f$ (interparticle grain boundary), and $\sim 0$ as $\phi_f \to 1$ (monolithic phase). n is set equal to 2 in equation (6). The choice of $3 > n > 1$ is found to give essentially the same end results. Following a similar discussion, the thermal resistance parameter $\alpha(\phi)$ is given by the heuristic expression $\alpha(\phi)=\alpha_{hp}(\phi)(1-\exp(-(t(\phi)/\lambda_h)^2))+\alpha_{pp}\exp(-(t(\phi)/\lambda_h)^2)$ so that $\alpha(\phi) \approx \alpha_{hp}(\phi)=R(\phi)k(\phi)/(a/2)$ for small $\phi$ and $\alpha_{pp} \approx R_p k_p/(a/2)$ as $\phi \to 1$, where $R(\phi) \approx 8/(C(\phi)v(\phi))$ and $R_p \approx 8/(Cv)_p$.[8, 19] In the case of hard-particle scattering, $F(\phi)$ in equation (6) becomes $(3\phi/2)(1-\exp(-(t(\phi)/\lambda_h)^2))$. Within the Debye model, at the high temperatures ($T > \theta_D$) of interest for higher ZT, phonons have a characteristic wavelength $\lambda_h \sim 3hv_h/2k\theta_D$, where $\theta_D$ is the Debye temperature and $v_h$ is the sound velocity.

Using Si and SiGe as core materials, we envision producing core-shell nanocomposites via in-situ intergranular phase segregation in nano-bulk Si and SiGe alloyed with immiscible elements, direct surface oxidation of Si and SiGe nanoparticles, or one of the chemical methods that are used to synthesize core-shell semiconductor quantum dots.[23] Suitably doped oxide, phosphide, and telluride phases are candidate shell materials. The materials used for the core and shell can also be reversed. The possibilities are numerous. Using yttrium-stabilized zirconia (YSZ) with large Seebeck coefficient ~400 μV/K as an example, the effective thermal conductivities of two n-type core-shell nanocomposites Si/YSZ and SiGe/YSZ, modeled as Si and SiGe nanoparticles embedded in YSZ, are calculated by the DEM method. Values of $k_{ho}$, $k_{po}$, $\lambda_h$, and sound velocities are given in Table I.[12, 19, 24-28] $C(\phi)$, $v(\phi)$, and $\alpha_{hp}(\phi)$ are obtained using the method described in reference 19. Equation (5) is numerically solved in conjunction with $F(\phi)$ in equation (6). As the volume fraction of the embedded particles increases the interparticle distance t shrinks, and phonon scattering in the host becomes more confined. Plots of lattice thermal conductivity for the two CSN systems as a function of $\phi$ and t (2×shell thickness) are presented in Figures 1-4. First, the model is validated by examining the monolithic nanomaterials. For the latter, k of the mono-grain-size phase can be computed by setting $F(\phi_f)=0$ at $\phi_f=1$. Noting that the derivative $dk/d\phi$ expressed by equation (5) must be continuous as $\phi \to 1$, a thermal conductivity of $k_{po}/(1+L_p/a+\alpha_{pp})$ is obtained. The same result holds in the hard-particle model. Comparison with experiment is shown in Table I. Despite the lack of further detailed information about the microstructure of nano-bulk Si and SiGe alloy,[12,15,29-31] the measured k values, k(expt.), of the two nanophases with the reported grain size ranges fall systematically within the ranges of computed k, k(calc.), for the same grain size ranges. Further validation of the present model is performed by examining the effect of large grain size range on k in nano-bulk Si. Reference 12 reported a nanostructure consisting of 50- to 100-nm grains in which many 10-nm nanocrystalline domains were also observed. The thermal conductivity was ~3.3 W/m/K. We obtained k=2.3 and 13.8 W/m/K for nano-bulk Si using single nanograin sizes of 10 and 100 nm, and k=2.3 and 4.26 W/m/K using single nanograin sizes of 10 and 20 nm. To make contact with the reported microstructure, we computed k(ϕ) for a silicon nanoparticles-in-matrix system in which ϕ volume fraction of 10-nm Si nanoparticles is embedded in a 100-nm grained Si-matrix. Good agreement with experiment was obtained near ϕ=0.3, which could be the volume fraction of 10-nm nanodomains that existed in the reported sample. In contrast, ab-initio calculation showed good agreement with experiment if a single grain size of 10 nm for the Si-matrix was assumed.[12] Overall, the good agreement between DEM model calculation and experiment obtained using only average phonon properties is remarkable. This finding may be attributed to the fact that phonon scattering in a disordered nanocomposite structure tend to mix the phonon states.

The influence of equation (6) on the thermal conductivity is evident by comparing the plots shown in Figures 1-4 for core sizes of 10, 20, 50, and 100 nm obtained using the core-shell model and hard-particle model. Several points stand out in the analysis. First of all, prominent depression of the CSN's lattice thermal conductivity below that obtained for the hard-particle model due to interparticle boundary scattering at increasing $\phi$ is noted in Figures 1 and 3. The thermal conductivities of both CSN-SiGe and Si are reduced significantly below those of nano-bulk SiGe and Si. Furthermore, although the YSZ host has a lower thermal conductivity than nano-bulk Si and also SiGe with grain size larger than 20 nm, all the plots for CSN-SiGe and two of the plots for CSN-Si are seen to dip below the thermal conductivity of YSZ. This can be seen more clearly in the k versus t plots shown in Figures 2 and 4. The above features are also present if the core and shell roles of YSZ and Si or SiGe are reversed. State-of-the-art core-shells have shell thickness of ~1-2 nm,[23] corresponding to t~2-4 nm. Accordingly, k(CSN) is computed for t=2 nm and 4 nm. k(CSN) values averaged over the core size range 10-20 nm for each t value are shown in Table I. Compared with nano-bulk Si and SiGe, up to 50% and 25% reductions in thermal conductivity are seen in CSN-Si and SiGe, respectively. Measured ZT, ZT(expt.), for nano-bulk Si[12] and SiGe[29, 30] were reported to be 0.7 and ~1.35 (1.3-1.4) at 1130 K, respectively. Without considering the possible beneficial impact on the power factor due to charge trapping at the core-shell interface, the ZT values of CSN, obtained from ZT(CSN)=(k(expt.)/k(CSN))*ZT(expt.), for the two shell thicknesses are found to increase quite significantly. In particular, ZT of 1.2 and 1.4 obtained for CSN-Si are nearly twice that of state-of-the-art nano-bulk Si, an inexpensive and abundant material for high-performance thermoelectric devices. By reducing the core size to 5-10 nm or interparticle distance to ~1 nm in both CSN systems, a high ZT of ~2 could be attained. Investigation of other core-shell nanocomposites using the present computational approach could lead to the discovery of other low thermal conductivity nanocomposites. From the experimental perspective, the fact that the core and shell can be independently tuned will enhance the search for higher ZT in cores-shell nano-bulk materials.

The differential effective medium method, which inherently included multiparticle effect, was employed to calculate the lattice thermal conductivity of core-shell based nanocomposites. In order to do so, the differential effective medium theory must be approached from a more physical viewpoint. Interparticle boundary scattering of phonons was found to contribute significantly to the reduction of thermal conductivity for Si and SiGe based core-shell nanocomposites, potentially achieving high thermoelectric ZT above that of monolithic nano-bulk phases.

**ACKNOWLEDGEMENTS**

The work is supported by a Department of Energy STTR Phase-2 Grant DE-SC0004317.

Table I. Parameters used in the calculation as well as calculated and measured lattice thermal conductivities at ~1130 K in the region of highest ZT. k(expt.) and k(calc.) are for nano-bulk Si and SiGe with grain size ranges in parentheses.. k(CSN) and ZT(CSN) are predicted averaged values for Si/YSZ and SiGe/YSZ core-shell composites with 10-nm and 20-nm core sizes, and for shell widths of 2 and 4 nm.

| Core/shell materials | $\theta_D$ (K) | $V_s$ (km/s) | $\lambda_h$ (nm) | $k_{h0}, k_{p0}$ (W/m/K) | k(expt.) (W/m/K) | k(calc.) (W/m/K) | k(CSN) (W/m/K) | ZT(CNS) |
|---|---|---|---|---|---|---|---|---|
| YSZ[24-26] | 527 | 3.7 | 0.51 | 1.7 | ---- | ---- | ---- | ---- |
| Si[12,19,27] | 645 | 6.0 | 0.66 | 31.2 | 3.3[12] (10-100nm) ~13[15]* (75-150nm) | 2.3-13.8 (10-100nm) 12.1-17.4 (75-150nm) | 1.66 (t=2nm) 1.86 (t=4nm) | 1.2, 1.4 |
| $Si_{80}Ge_{20}$[19,27,28] | 587 | 5.2 | 0.63 | 2.6 | 1.35[29,30] (10-20nm) 1.82[31] (20-50nm) | 1.20-1.63 (10-20 nm) 1.20-2.10 (20-50nm) | 1.0 (t=2nm) 1.25 (t=4nm) | 1.45, 1.8 |

* Extrapolated value.

Figure captions.

Figure 1. Lattice thermal conductivity of $Si_{80}Ge_{20}$/YSZ core-shell nanocomposites as a function of volume fraction of $Si_{80}Ge_{20}$ core particles at T~1130 K. Dashed and solid curves are obtained using the hard-particle model (HP-model) and F($\phi$) function from equation (6) (CS-model), respectively. Core particle size (from top to bottom): 100, 50, 20, and 10 nm.

Figure 2. Lattice thermal conductivity of $Si_{80}Ge_{20}$/YSZ core-shell nanocomposites as a function of interparticle distance (shortest distance between particles, or twice the shell width) between $Si_{80}Ge_{20}$ core particles at T~1130 K. Dashed and solid curves are obtained using the hard-particle model (HP-model) and F($\phi$) function from equation (6) (CS-model), respectively. Core particle size (from top to bottom): 100, 50, 20, and 10 nm.

Figure 3. Lattice thermal conductivity of Si/YSZ core-shell nanocomposites as a function of volume fraction of Si core particles at T~1130 K. Dashed and solid curves are obtained using the hard-particle model (HP-model) and F($\phi$) function from equation (6) (CS-model), respectively. Core particle size (from top to bottom): 150, 100, 75, and 10 nm.

Figure 4. Lattice thermal conductivity of Si/YSZ core-shell nanocomposites as a function of interparticle distance (shortest distance between particles, or twice the shell width) between Si core particles at T~1130 K. Dashed and solid curves are obtained using the hard-particle model (HP-model) and F($\phi$) function from equation (6) (CS-model), respectively. Core particle size (from top to bottom): 150, 100, 75, and 10 nm.

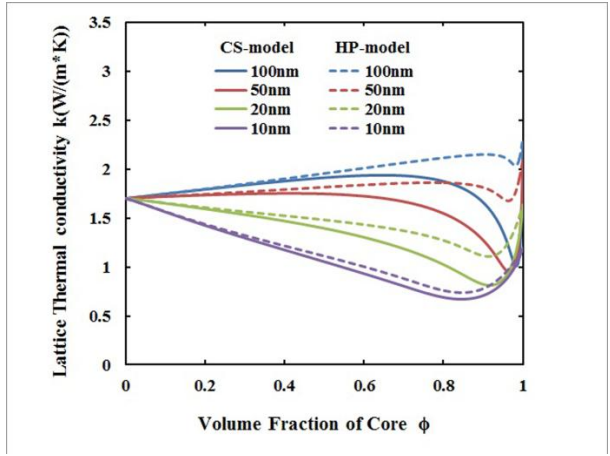

Figure 1

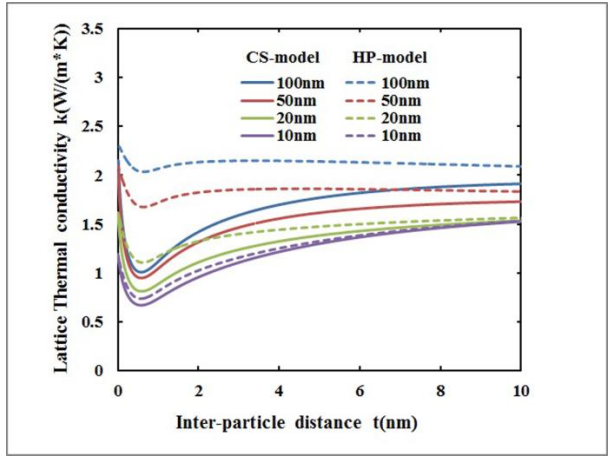

Figure 2

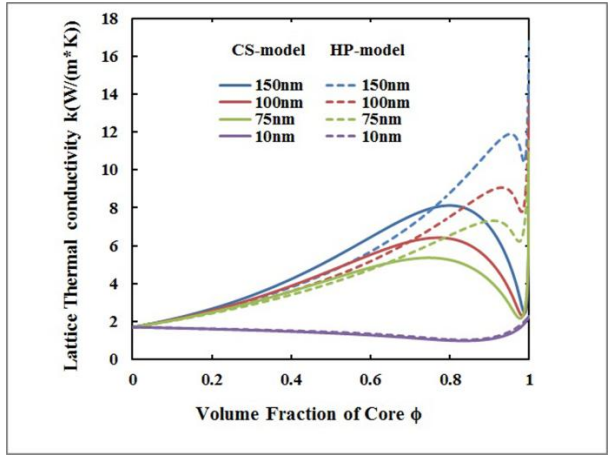

Figure 3

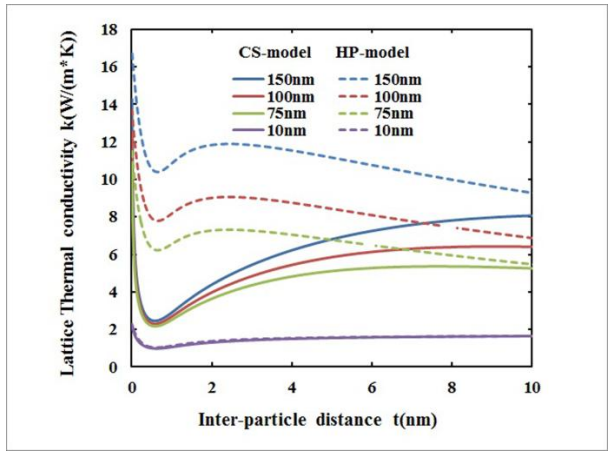

Figure 4